\documentclass[conf]{new-aiaa}
\usepackage[utf8]{inputenc}
\usepackage{bm}
\usepackage{graphicx}
\usepackage{amsmath}
\usepackage[version=4]{mhchem}
\usepackage{siunitx}
\usepackage{longtable,tabularx}
\usepackage{float}

\title{Reinforcement Learning for Angle-Only Intercept Guidance of Maneuvering Targets}

\author{Brian Gaudet\footnote{Co-Founder, DeepAnalytX LLC, Affiliated Engineer, Department of Aerospace and Mechanical Engineering} and Roberto Furfaro\footnote{Professor, Department of Systems and Industrial Engineering, Department of Aerospace and Mechanical Engineering}}
\affil{University of Arizona, 1127 E. Roger Way, Tucson Arizona, 85721}
\author{Richard Linares\footnote{Charles Stark Draper Assistant Professor, Department of Aeronautics and Astronautics, Senior Member AIAA, E-mail:linaresr@mit.edu}}
\affil{Massachusetts Institute of Technology, Cambridge, MA 02139}

\begin{document}

\maketitle

\begin{abstract}
 We present a novel guidance law that uses observations consisting solely of seeker line of sight angle measurements and their rate of change. The policy is optimized using reinforcement meta-learning and demonstrated in a simulated  terminal phase  of a mid-course exo-atmospheric interception.  Importantly, the guidance law does not require range estimation, making it particularly suitable for passive seekers. The optimized policy maps stabilized seeker line of sight angles and their rate of change directly to commanded thrust for the missile's divert thrusters. The use of reinforcement meta-learning allows the optimized policy to adapt to target acceleration, and we demonstrate that the policy performs as well as augmented zero-effort miss guidance with perfect target acceleration knowledge.   The optimized policy is computationally efficient and requires minimal memory, and should be compatible with today's flight processors.  
 
\end{abstract}

\section{Introduction}
\lettrine{E}{xo-atmospheric} interception of ballistic targets is particularly challenging due to the hit to kill requirement and relatively small size of a ballistic re-entry vehicle (BRV), typically 45 to 60 centimeters in diameter.  Successful interception requires both a small miss distance and a suitable impact angle, with miss distance requirements of 50 cm implied by the BRV and missile dimensions. Moreover, the missile must  autonomously discriminate between threats and decoys. The interception problem is significantly complicated by warheads with limited maneuvering capability. Both spiral and bang-bang maneuvers could potentially be executed by a BRV without compromising the BRV's accuracy. These maneuvers could be executed either in response to the BRV's sensor input (if so equipped) or periodically executed during the portion of the trajectory where interception is likely.  Another complication of exo-atmospheric interception is that  the high altitude requires the use of divert thrusters rather than control surfaces, with current implementation using pulsed divert  thrusters. As the missile burns fuel, its center of mass shifts, and the divert thrusts cause a tumbling motion that requires compensation from the attitude control thrusters. Fuel efficiency is also critical, as the missile loses all control authority when its fuel is depleted.

Recent work in exo-atmospheric guidance law development include \cite{ratnoo2009collision}, where a collision geometry based guidance law is developed that attempts to keep the missile on a collision triangle with the target using range and angle measurements. The authors demonstrate improved capture range and miss distance for the case of a non-maneuvering target as compared to a zero effort miss guidance law.  In \cite{gutman2019exoatmospheric}, the authors develop a guidance law suitable for exo-atmospheric intercepts using linear quadratic optimization; importantly the guidance law requires an estimate of the relative range and velocity vectors and assumes a non-maneuvering target. In \cite{zarchan2012tactical:1}, Zarchan develops a pulsed guidance law that removes the zero effort miss during the homing phase by precomputing the required number of pulses and setting each pulse width based off of the current time to go, closing velocity, and line-of-sight rate of change. 

The guidance laws described above require, at a minimum, the measurement of range (which allows the estimation of closing velocity), with some requiring estimation of the relative positions and velocities. Active sensors such as radar can provide both range and closing velocity measurements, and by including angle measurements,  a Kalman filter \cite{kalman1961new} can estimate relative positions, velocities, and target acceleration. On the other hand, passive electro-optical IR seekers such as those used in the Navy's SM-3 exo-atmospheric missile \cite{sullins2001exo} do not provide range measurements.  Note that it is  possible to estimate range at lock-on using the target's radiant intensity and an estimate of the attenuation of the signal with distance \cite{siouris2004missile}, after which the change in range could potentially be estimated from integrated IMU measurements. However, it seems likely that using measured radiant intensity could be easily fooled by countermeasures. Finally, we note that the well-known proportional navigation guidance law can be implemented without range information \cite{shneydor1998missile} $\mathbf{a}_\mathrm{M}=N\boldsymbol{\omega}\mathbf{v}_\mathrm{M}$, where $\boldsymbol{\omega}$ is the rate of change of the line-of-sight vector, $\mathbf{a}_\mathrm{M}$ the commanded missile acceleration, and $\mathbf{v}_\mathrm{M}$ an estimate of the missile's velocity vector. However, an accurate estimation of $\mathbf{v}_\mathrm{M}$ is probably a more difficult problem than the estimation of range and range rate from angle-only measurements.  

The work relating to a missile guidance law using angle-only measurements is relatively scarce. In \cite{taur1999passive} the authors develop an extended Kalman filter that can estimate range from angle measurements. However, the experimental results are for endo-atmospheric intercepts, and in some scenarios it takes several seconds for the filter to converge, while in other engagement scenarios the filter does not converge to an accurate estimate. The authors also found that it was not possible to reliably estimate acceleration, which is required for advanced guidance laws such as  augmented  proportional navigation. Interestingly, the authors enhance observability by creating a guidance law that is a function of both the seeker angles and their rate of change. In this work we also use both seeker angles and their rate of change, but do not attempt to estimate range and closing velocity. In \cite{song1996practical} an augmented proportional navigation guidance law is modified to induce rotations in the line-of-sight angles during the engagement,  allowing a modified gain extended Kalman filter to estimate the full planar engagement state from angle-only measurements.  In the simulation results presented, the rotations do not impact performance, but that may not be the case in general. The method of inducing line-of-sight rotations was used more recently in \cite{reisner2013optimal} to estimate the engagement state from angle-only measurements. Importantly, the results were compared to the Proportional Navigation (PN) guidance law that was not augmented with a target acceleration estimate. In contrast, we compare our guidance law's performance to augmented PN with perfect knowledge of the engagement state. Finally, in \cite{kim2018look}, the authors develop a guidance law using sliding mode theory that is effective against a stationary target and implements impact angle and field of view constraints. 

Regardless of seeker type, a system that can map seeker angle measurements directly to actuator commands has several potential advantages. First, the state estimation problem is simplified, as only the seeker angles need to be estimated. With a fast enough sensor measurement frequency, this could potentially be accomplished by simply averaging the angle measurements between guidance system updates. Second, there is less potential for problems arising from different estimation biases for relative position, velocity, and target acceleration. Third, with less processing requirements, the guidance frequency can be increased. Finally, using only angle measurements may be less susceptible to countermeasures such as range gate pull-off \cite{desk1997electronic}, where the target generates a radar signal that forces the radar's tracking gates to be pulled away from the target echo.   It seems intuitive that a guidance system that does not use range measurements would be immune to such countermeasures. Although there are many deployed missiles with guidance systems compatible with passive seekers, to the best of our knowledge, there is currently no method to apply the optimal control framework to optimizing such a guidance law using angle-only measurements, or a robust, fast-converging method to estimate the full engagement state solely from angle measurements without inducing line-of-sight rotation during the engagement. It is possible that the ability to introduce an explicit cost function (rewards in the RL framework) and constraints can boost performance beyond what is possible today with passive seekers. The purpose of this paper to take the first step towards that goal.

In this work we optimize a guidance law using  reinforcement learning (RL)\footnote{See Section \ref{RL} for an overview of RL}. The optimized policy maps estimates of the two seeker angles and their rate of change directly to divert thrust commands, and respects seeker field of view and maximum thrust constraints. The policy is learned by having an agent instantiating the policy interact episodically with an environment. Each episode consists of an engagement scenario with randomized parameters.  Here we consider a maneuverable ballistic re-entry vehicle high-altitude interception scenario, where the intercepting missile must destroy the target kinetically via a direct hit (miss less than 50 cm). The engagement scenario is considerably simplified from a realistic engagement. First, we only simulate the interception's terminal phase.  Second, we do not generate realistic ballistic trajectories for the missile or target, and neglect the force of gravity. Note that it is a common practice to neglect gravity when initially developing a new guidance law [\citenum{ratnoo2009collision}, \citenum{zarchan2012tactical:1}, \citenum{song1996practical}, \citenum{taur1999passive}]. Third, we do not address target discrimination or attitude control. The engagement scenario is then a simple head on engagement with the target having the speed advantage, with initial missile and target speeds of 3000 m/s and 4000 m/s respectively. The target executes a randomized bang-bang maneuver during the intercept. This is a realistic maneuver for a re-entry vehicle to execute in order to evade interception, as it does not dramatically modify the re-entry vehicles trajectory. We use a 2:1 ratio of missile to target thrust capability.  The details of this engagement scenario are provided in Section \ref{engagement}.

The optimized policy is then then tested, and performance is compared to an augmented zero-effort-miss (ZEM) policy \cite{zarchan2012tactical:2} using the full ground-truth engagement state (relative position and velocity, and target acceleration). Since the engagement parameters are randomized for each episode, the test engagements are novel in that the engagement parameters were not experienced by the agent during optimization. We find that the RL policy has slightly better accuracy, but at the expense of slightly increased average fuel consumption.  We attribute the RL policy's performance to the use of a recurrent neural network layer in the policy and value function, which allows the policy to adapt to a particular target maneuver in real time. Specifically, the recurrent layer's hidden state evolves differently in response to target maneuvers in a particular engagement, allowing the mapped actions to take into account the maneuver. As opposed to augmented ZEM, where the state estimation filter must estimate an acceleration, the RL policy instead adapts to the target maneuver in real time.  In previous work regarding an adaptive policy for asteroid close proximity maneuvers that could adapt to unknown environments and actuator failures \cite{gaudet2019adaptive}, we also found that using a recurrent layer in the policy and value function boosted performance. Mapping observations to actions requires only four small matrix multiplications, which takes less than 1 ms on a 2.3 Ghz CPU and requires around 64KB of memory. In this work we use a 100 ms guidance cycle, and we expect that the policy would easily run on modern flight computers.

\section{Problem Formulation}

\subsection{Missile Configuration}

The missile is modeled as a cylinder of height  1 m and radius 0.25 m about the missile's body frame x-axis  with a wet and dry mass of 50 kg and 25 kg, respectively. Four divert thrusters  thrusters are positioned as shown in Table \ref{tab:thrusters}. We assume that attitude control thrusters keep the missile's attitude constant during the engagement, but do not address an implementation of an attitude control policy in this work.  The divert thrusters can be switched on or off at the guidance frequency of 10 Hz. 

\begin{table}[h]
	\fontsize{10}{10}\selectfont
    \caption{Body Frame Thruster Locations.}
   \label{tab:thrusters}
        \centering 
   \begin{tabular}{c |r | r | r | r | r | r } % Column formatting, 
      \hline
      & \multicolumn{3}{r}{Divert Thrust Direction Vector}\vline & \multicolumn{3}{r}{Location}  \\
       \hline
      Thruster & x  & y  & z  &  x (m) & y (m) & z (m)\\
      \hline
      1 & 0.00 & -1.00 & 0.00 & 0.00 & -0.25 & 0.00    \\
      2 & 0.00 & 1.00 & 0.00 & 0.00 & 0.25 & 0.00  \\
      3 & 0.00 & 0.00 & 1.00 & 0.00 & 0.00 & 0.25 \\
      4 & 0.00 & 0.00 & -1.00 & 0.00 & 0.00 & -0.25  \\
      
   \end{tabular}
\end{table}

We assume that the seeker is perfectly stabilized and has perfect tracking capability, with a 135 degree field of view. At the start of the homing phase, the seeker's attitude is set to the missile body frame attitude, where it remains fixed (stabilized) during the homing phase. To be clear, in the absence of any change in the missile's attitude during the homing phase, this implies that the seeker and body reference frames remain aligned, i.e., they both have the same attitude with respect to the inertial frame. However if the missile's attitude changes during the engagement, the seeker's attitude remains at the missile body frame attitude measured at the start of the engagement. This stabilization insures that missile attitude changes are not interpreted as target maneuvers during the engagement.

As the seeker's sensor tracks the target from this stabilized reference frame, we can define the angles between the seeker boresight axis and the seeker reference frame $y$ and $z$ axes as the seeker angles $\theta_u$ and $\theta_v$.  Further, define $\mathbf{C}_\mathrm{SN}(\mathbf{q}_0)$ as the direction cosine matrix (DCM) mapping from the intertial frame to the stabilized seeker platform reference frame, with $\mathbf{q}_0$ being the missile's attitude at the start of the homing phase. We can now transform  the target's relative position in the inertial reference frame $\mathbf{r}_\mathrm{TM}^\mathrm{N}$ into the seeker reference frame as shown in Eq.~\eqref{eq:seeker1}.

\begin{equation}
\label{eq:seeker1}
\mathbf{r}_\mathrm{TM}^\mathrm{S} = [\mathbf{C}_\mathrm{SN}(\mathbf{q}_0)]\mathbf{r}_\mathrm{TM}^\mathrm{N}
\end{equation}

Defining the line-of-sight unit vector in the seeker reference frame as $\boldsymbol{\hat\lambda}^\mathrm{S} = \dfrac{\mathbf{r}_\mathrm{TM}^\mathrm{S}}{\|\mathbf{r}_\mathrm{TM}\|^\mathrm{S}}$ and the seeker frame unit vectors $\hat{u}=\begin{bmatrix} 0 & 1 & 0 \end{bmatrix}$ , $\hat{v}=\begin{bmatrix} 0 & 0 & 1\end{bmatrix}$  we can then compute the seeker angles as the orthogonal projection of the seeker frame LOS vector onto $\hat u$ and $\hat v$ as shown in Eqs.~\eqref{eq:seeker3} and \eqref{eq:seeker4}.

\begin{subequations}
\begin{align}
\theta_{u} &= \mathrm{arcsin}(\boldsymbol{\hat\lambda}^\mathrm{S} \cdot \hat{u})\label{eq:seeker3}\\
\theta_{v} &= \mathrm{arcsin}(\boldsymbol{\hat\lambda}^\mathrm{S} \cdot \hat{v})\label{eq:seeker4}
\end{align}
\end{subequations}

The guidance policy described in the following will map these seeker angles $\theta_u$ and $\theta_v$ and their rate of change to divert thrust commands. 

\subsection{Engagement Scenario}\label{engagement}
\begin{figure}[h]
\begin{center}
\includegraphics[width=.6\linewidth]{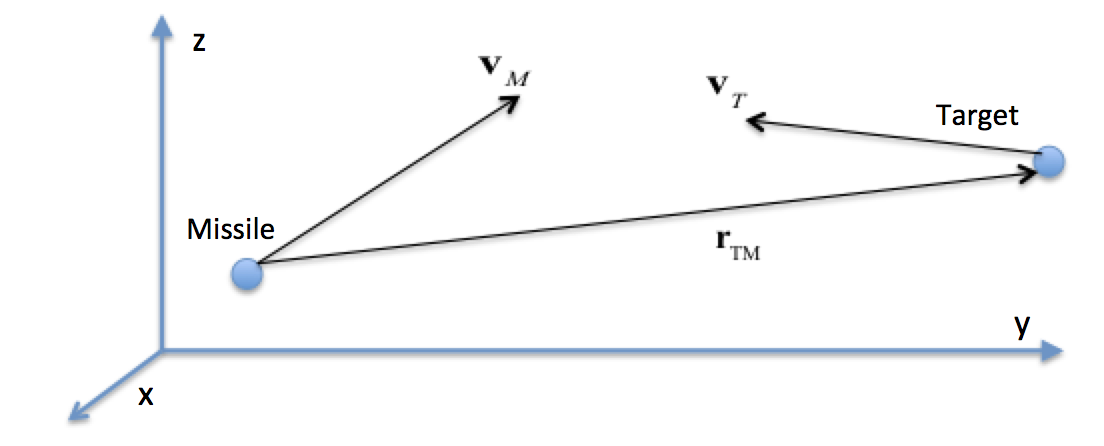}
\caption{Engagement}
\label{fig:engagement}
\end{center}
\end{figure}

In this work we use a simplified engagement scenario. Instead of modeling the missile and target trajectories using Lambert guidance to determine the homing phase initial conditions, the engagement is modeled as a simple head-on engagement. Referring to Fig. \ref{fig:engagement}\footnote{In this figure, the illustrated vectors are not in general within the y-z plane}, where the missile velocity vector, target velocity vector, and relative range vector are given as $\mathbf{v}_\mathrm{M}$, $\mathbf{v}_\mathrm{T}$, and $\mathbf{r_\mathrm{TM}}$, we can define the target's initial position $\mathbf{r}_\mathrm{T}$ in a missile centered reference frame in terms of the range vector from missile to target $\|\bf r_\mathrm{TM}\|$ and angles $\theta$ and $\phi$ as given in Equations \eqref{eq:eng1}, \eqref{eq:eng2}, and \eqref{eq:eng3}.

\begin{subequations}
\begin{align}
r_{\mathrm{T}_x} &= \|\bf r_\mathrm{TM}\|\mathrm{sin}(\theta)\mathrm{cos}(\phi)\label{eq:eng1}\\
r_{\mathrm{T}_y} &= \|\bf r_\mathrm{TM}\|\mathrm{sin}(\theta)\mathrm{sin}(\phi)\label{eq:eng2}\\
r_{\mathrm{T}_z} &= \|\bf r_\mathrm{TM}\|\mathrm{cos}(\theta)\label{eq:eng3}
\end{align}
\end{subequations}

Further, we can represent the target's initial velocity vector $\mathbf{v}_\mathrm{T}$ in terms of the magnitude of the target velocity $\|\mathbf{v}_\mathrm{T}\|$ and angles $\alpha$ and $\beta$ as shown in Equations \eqref{eq:he1}, \eqref{eq:he2}, and \eqref{eq:he3}.

\begin{subequations}
\begin{align}
v_{\mathrm{T}_x} &= \|\mathbf{v}_\mathrm{T}\|\mathrm{sin}(\beta)\mathrm{cos}(\alpha)\label{eq:he1}\\
v_{\mathrm{T}_y} &= \|\mathbf{v}_\mathrm{T}\|\mathrm{sin}(\beta)\mathrm{sin}(\alpha)\label{eq:he2}\\
v_{\mathrm{T}_z} &= \|\mathbf{v}_\mathrm{T}\|\mathrm{cos}(\beta)\label{eq:he3}
\end{align}
\end{subequations}

\begin{figure}[h]
\begin{center}
\includegraphics[width=.6\linewidth]{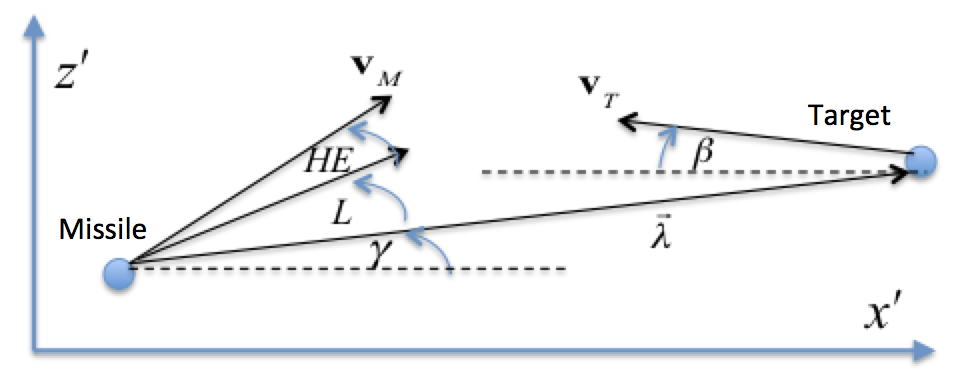}
\caption{Planar Heading Error}
\label{fig:HE}
\end{center}
\end{figure}

A collision triangle can be defined in a plane that is not in general aligned with the coordinate frame shown in Fig. \ref{fig:engagement}, and is illustrated in Fig. \ref{fig:HE}.  Here we define the required lead angle $L$ for the missile's velocity vector $\mathbf{v}_m$ as the angle that will put the missile on a collision triangle with the target in terms of the target velocity $\mathbf{v}_\mathrm{T}$, line-of-sight angle $\gamma$, and the magnitude of the missile velocity as shown in Eqs.~\eqref{eq:lead1} through \eqref{eq:lead3}.

\begin{subequations}
\begin{align}
    L &= \mathrm{arcsin}\left(\frac{\|\mathbf{v}_\mathrm{T}\|\mathrm{sin}(\beta+\gamma)}{\|\mathbf{v}_\mathrm{M}\|}\right)\label{eq:lead1}\\
    v_{m_y} &= \|\mathbf{v}_{\mathrm{M}}\|\mathrm{cos}(L + \gamma)\label{eq:lead2}\\
    v_{m_z} &= \|\mathbf{v}_{\mathrm{M}}\|\mathrm{sin}(L + \gamma)\label{eq:lead3}\\
\end{align}
\end{subequations}

This formulation is easily extended to a three dimensional engagement by defining a plane normal as $\mathbf{\hat{v}}_t \times \boldsymbol{\hat{\lambda}}$, rotating $\mathbf{v}_t$ and $\boldsymbol{\hat{\lambda}}$ onto the plane, calculating the required planar missile velocity, and then rotating this back into the original reference frame. Thus in $\mathbb{R}^3$ we define a heading error (HE) as the the angle between the missile's initial velocity vector and the velocity vector associated with the lead angle required to put the missile on a collision heading with the target. We also define the initial attitude error as the angle between the missile's velocity vector and the body frame x-axis at the start of the engagement.

\begin{figure}[h]
\begin{center}
\includegraphics[width=.8\linewidth]{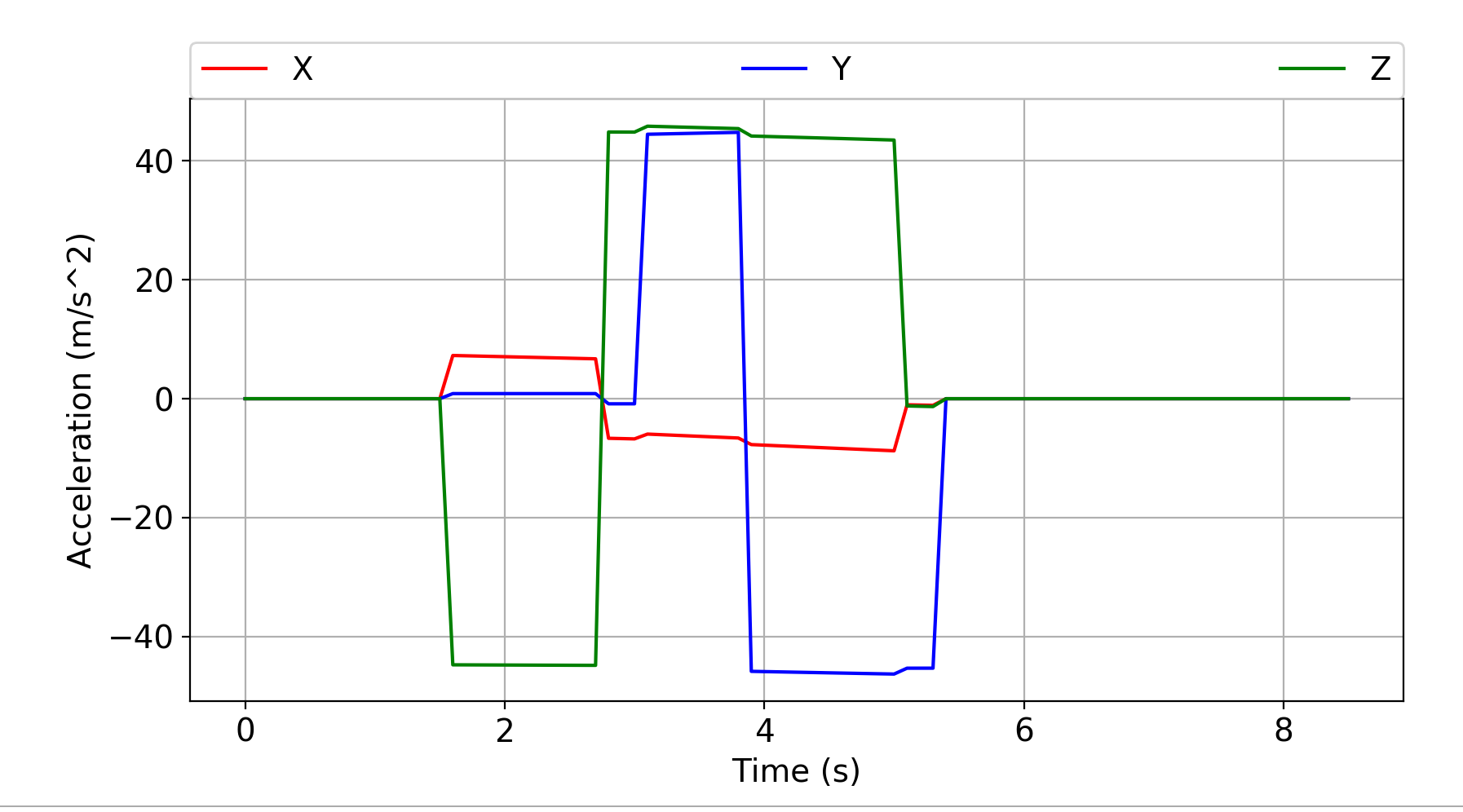}
\caption{Sample Target Maneuver}
\label{fig:TM}
\end{center}
\end{figure}

We consider a bang-bang target maneuver, where the acceleration is applied orthogonal to the target's velocity vector. The maneuver has varying acceleration levels up to a maximum of $\mathrm{5*9.81\ m/s}^2$, and with random start time, duration, and switching time. A sample target maneuver is shown Fig. \ref{fig:TM}, note that in some cases the maneuver duration is considerably longer, with a complete cycle extending beyond the time of intercept.

We can now list the range of engagement scenario parameters in Table \ref{tab:ic}.  During optimization and testing, these parameters are drawn uniformly between their minimum and maximum values.  The generation of heading error is handled as follows. We first calculate the optimal missile velocity vector that puts the missile on a collision triangle with the target as described previously. Treating this velocity vector as the axis of a cone, the heading error defined in $\mathbb{R}^3$ is then $\theta_\mathrm{HE}=\mathrm{atan}\frac{r}{h}$, where $r$ and $h$ are the cone's radius and height. Similarly, we define an ideal initial attitude as having the missile's x-axis aligned with its velocity vector, and use a similar method to perturb this by the attitude error such that angle between the missile's ideal and actual velocity vector is equal the the attitude error.

\begin{table}[H]
	\fontsize{10}{10}\selectfont
    \caption{Initial Conditions}
   \label{tab:ic}
        \centering 
   \begin{tabular}{l | r | r } % Column formatting, 
      \hline
      Parameter & min & max \\
      \hline
      Range $\|\mathbf{r}_\mathrm{TM}\|$ (km) & 50 & 55\\
      Missile Velocity Magnitude (m/s) & 3000 & 3000 \\
      Target Position angle $\theta$ (degrees) & -10 & 10 \\
      Target Position angle $\phi$ (degrees) & -10 & 10 \\
      Target Velocity Magnitude (m/s) & 4000 & 4000 \\
      Target Velocity angle $\beta$ (degrees) & -10 & 10 \\
      Target Velocity angle $\alpha$ (degrees) & -10 & 10 \\
      Heading Error (degrees) & 0 & 5 \\
      Attitude Error (degrees) & 0 & 5 \\
      Target Acceleration  $\mathrm{m/s^2}$ & -5*9.81 & 5*9.81
   \end{tabular}
\end{table}

\subsection{Equations of Motion}

The force $\mathbf{F}_{B}$ and torque $\mathbf{L}_{B}$ in the missile's body frame for a given commanded thrust depends on the placement of the thrusters in the missile structure. We can describe the placement of each thruster through a body-frame direction vector $\mathbf{d}$ and position vector $\mathbf{r}$, both in $\mathbb{R}^3$.  The direction vector is a unit vector giving the direction of the body frame force that results when the thruster is fired.  The position vector gives the body frame location with respect to the missile centroid,   where the force resulting from the thruster firing is applied for purposes of computing torque, and in general the center of mass ($\mathbf{r}_\mathrm{com}$) varies with time as fuel is consumed. For a missile with $k$ thrusters, the body frame force and torque associated with one or more  thrusters firing is then as shown in Eqs. \eqref{eq:Thruster_modela} and \eqref{eq:Thruster_modelb}, where $T_{cmd_{i}}\in[T_{min},T_{max}]$ is the commanded thrust for thruster $i$, $T_{min}$ and $T_{max}$ are a thruster's minimum and maximum thrust, $\mathbf{d}^{(i)}$ the direction vector for thruster $i$, and $\mathbf{r}^{(i)}$ the position of thruster $i$. The total body frame force and torque are calculated by summing the individual forces and torques.

\begin{subequations}
\begin{align}
	{\mathbf{F}_{B}}&={\sum_{i=1}^{k}\mathbf{d}^{(i)} T_{cmd}^{(i)}}\label{eq:Thruster_modela}\\
	{\mathbf{L}_{B}}&={\sum_{i=1}^{k}(\mathbf{r}^{(i)}-\mathbf{r}_\mathrm{com})\times\mathbf{F}_{B}^{(i)}}\label{eq:Thruster_modelb}
\end{align}
\end{subequations}

The dynamics model uses the missile's current attitude $\mathbf{q}$ to convert the body frame thrust vector to the inertial frame as shown in in Eq. \eqref{eq:BtoN} where $[\mathbf{BN}](\mathbf{q})$ is the direction cosine matrix mapping the inertial frame to body frame obtained from the current attitude parameter $\mathbf{q}$.

\begin{equation}
	\label{eq:BtoN}
	\mathbf{F}_{N}=\left[\left[\mathbf{BN}\right](\mathbf{q})\right]^{T}\mathbf{F}_{B}
\end{equation}

The missile's translational motion is modeled as shown in \ref{eq:EQOMa} through \ref{eq:EQOMc}. Since we are not accurately modeling the initial conditions as a ballistic intercept (i.e., using Lambert guidance), we do not model the gravitational acceleration.

\begin{subequations}
\begin{align}
	{\Dot{\mathbf r}} &= {{\mathbf v}}\label{eq:EQOMa}\\
	{\Dot{\bf v}} &= \frac{{{\bf F}_{N}}}{m}\label{eq:EQOMb}\\
	\Dot{m} &= -\frac{\sum_{i}^{k}\lVert{{\bf F}_{B}}^{(i)}\rVert}{I_\text{sp}g_\text{ref}} \label{eq:EQOMc}
\end{align}
\end{subequations}
Here  ${{\bf F}_{N}}^{(i)}$ is the inertial frame force as given in Eq.~\eqref{eq:BtoN}, $k$ is the number of thrusters, $g_\text{ref}=9.8$ $\text{m}/\text{s}^{2}$,  $\mathbf{r}$ is the missile's position in the engagement reference frame

The target is modeled as shown in Eqs.~\eqref{eq:TEQOMa} and \eqref{eq:TEQOMb}, where $\mathbf{a}_{\mathrm{T}_\mathrm{com}}$ is the commanded acceleration for the target maneuver.

\begin{subequations}
\begin{align}
	{\Dot{\mathbf r}} &= {{\mathbf v}}\label{eq:TEQOMa}\\
	{\Dot{\bf v}} &= \mathbf{a}_{\mathrm{T}_\mathrm{com}}\label{eq:TEQOMb}
\end{align}
\end{subequations}

The equations of motion are updated using fourth order Runge-Kutta integration.  For ranges greater than 1000 m, a timestep of 20 ms is used, and for the final 1000 m of homing, a timestep of 0.067 ms is used in order to more accurately measure miss distance; this technique is borrowed from \cite{zarchan2012tactical:4}.  For the ZEM guidance law, this results in a 100\% hit rate (miss $<$ 50 cm) with no target maneuver and zero heading error at a guidance frequency of 10 Hz. We also checked that decreasing the integration step size to 10 ms did not improve performance for the augmented ZEM guidance law.

\section{Guidance Law Development}

\subsection{Reinforcement Learning Overview}\label{RL}

\begin{figure}[h]
\begin{center}
\includegraphics[width=.75\linewidth]{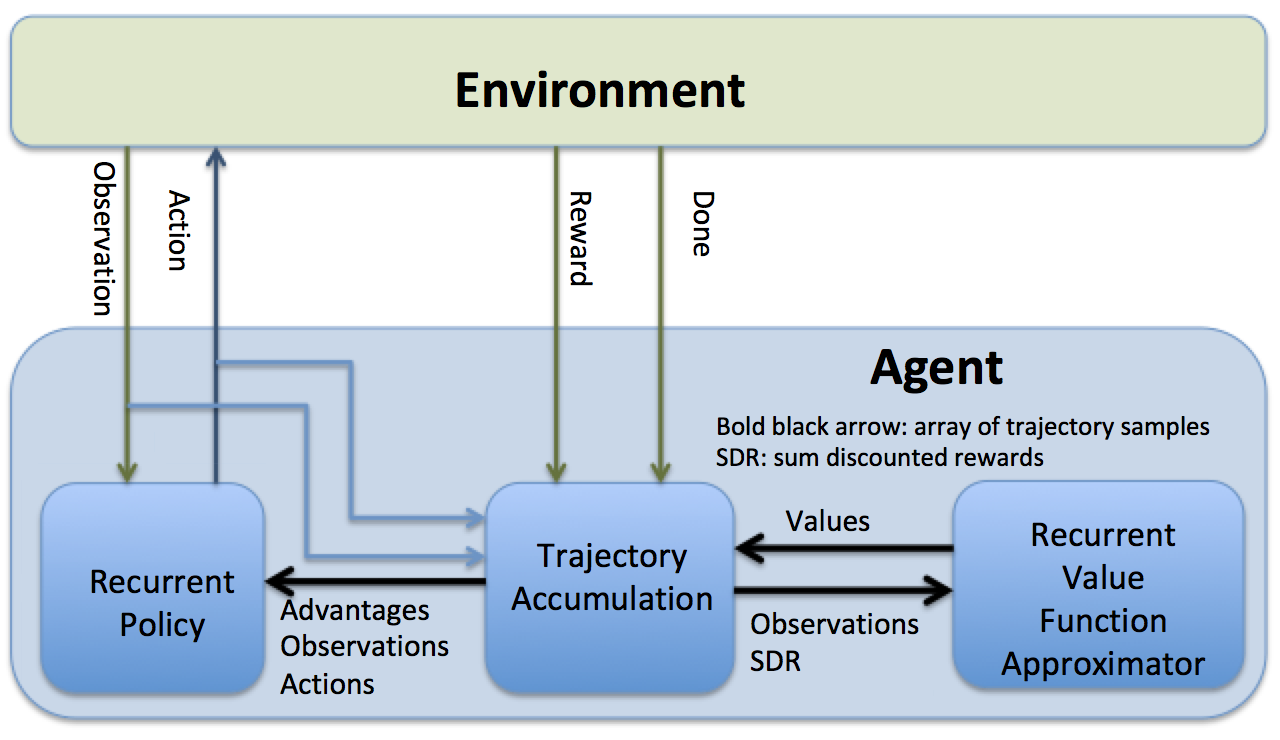}
\caption{Engagement}
\label{fig:Agent_env_detail}
\end{center}
\end{figure}

In the RL framework, an agent learns through episodic interaction with an environment how to successfully complete a task by learning a \textit{policy} that maps observations to actions. The environment initializes an episode by randomly generating a ground truth state, mapping this state to an observation, and passing the observation to the agent. These observations could be an estimate of the ground truth state from a Kalman filter or could be raw sensor outputs such as seeker angle measurements and radar range and closing velocity measurements, or a multi-channel pixel map from an electro-optical sensor.  The agent uses this observation to generate an action that is sent to the environment; the environment then uses the action and the current ground truth state to generate the next state and a scalar reward signal.  The reward and the observation corresponding to the next state are then passed to the agent. The process repeats until the environment terminates the episode, with the termination signaled to the agent via a done signal. Possible termination conditions include the agent completing the task, satisfying some condition on the ground truth state (such negative closing speed), or violating a constraint.  Typically, trajectories from some fixed number of episodes (referred to as rollouts) are collected during interaction between the agent and environment, and used to update the policy and value functions. The interface between agent and environment is depicted in Fig. \ref{fig:Agent_env_detail}.

A Markov Decision Process (MDP) is an abstraction of the environment, which in a continuous state and action space, can be represented by a state space $\mathcal{S}$, an action space $\mathcal{A}$, a state transition distribution $\mathcal{P}(\mathbf{x}_{t+1}|\mathbf{x}_t,\mathbf{u}_t)$, and a reward function $r=\mathcal{R}(\mathbf{x}_t,\mathbf{u}_t))$, where $\mathbf{x} \in \mathcal{S}$ and $\mathbf{u} \in \mathcal{A}$, and $r$ is a scalar reward signal. We can also define a partially observable MDP (POMDP), where the state $\mathbf{x}$ becomes a hidden state, generating an observation $\mathbf{o}$ using an observation function $\mathcal{O}(\mathbf{x})$ that maps states to observations. The POMDP formulation is useful when the observation consists of raw sensor outputs, as is the case in this work.  In the following, we will refer to both fully observable and partially observable environments as POMDPs, as an MDP can be considered a POMDP with an identity function mapping states to observations.

The agent operates within an  environment defined by the POMDP, generating some action $\mathbf{u}_t$ based off of the observation $\mathbf{o}_t$, and receiving reward $r_{t+1}$ and next observation $\mathbf{o}_{t+1}$. Optimization involves maximizing the sum of (potentially discounted) rewards over the trajectories induced by the interaction between the agent and environment. Constraints such as minimum and maximum thrust,  attitude compatible with sensor field of view, maximum angle of attack, and  maximum rotational velocity, can be included in the reward function, and will be accounted for when the policy is optimized. Note that there is no guarantee on the optimality of trajectories induced by the policy, although in practice it is possible to get close to optimal performance by tuning the reward function.

Reinforcement meta-learning differs from generic reinforcement learning in that the agent learns to quickly adapt to novel POMPDs by learning over a wide range of POMDPs. These POMDPs can include different environmental dynamics, actuator failure scenarios, mass and inertia tensor variation, and varying amounts of sensor distortion. Learning within the RL meta-learning framework results in an agent that can quickly adapt to novel POMDPs, often with just a few steps of interaction with the environment. There are multiple approaches to implementing meta-RL.  In \cite{finn2017model}, the authors design the objective function to explicitly make the model parameters transfer well to new tasks. In \cite{mishra2018simple}, the authors demonstrate state of the art performance using temporal convolutions with soft attention. And  in \cite{frans2017meta}, the authors use a hierarchy of policies to achieve meta-RL. In this proposal, we use a different approach \cite{wang2016learning} using a recurrent policy and value function. Note that it is possible to train over a wide range of POMDPs using a non-meta RL algorithm. Although such an approach typically results in a robust policy, the policy cannot adapt in real time to novel environments. To better understand the differences between RL and current practice using optimal control, a comparison of RL and optimal control approaches to guidance and control are given in Table~\ref{tab:rl_vs_oc}. The point of the comparison is not to make the case that one approach should be preferred over the other, but rather to suggest the scenarios where it might make sense to use the RL framework to solve guidance and control problems.

\begin{table}[h!]
	\fontsize{10}{10}\selectfont
    \caption{A Comparison of Optimal Control and RL \cite{tedrake2015underactuated}}
   \label{tab:rl_vs_oc}
        \centering 
   \renewcommand{\arraystretch}{1.5}
   \begin{tabular}{ p{5.5cm} | p{5.5cm} } % Column formatting, 
      \bf{Optimal Control} & \bf{Reinforcement Learning}\\
      \hline
      Single trajectory (except for trivial cases where HJB equations can be solved) & Global over theatre of operations \\
      Unbounded run time except for special cases such as convex constraints & Extremely fast run time for trained policy ($<$ 1ms in this work) \\
      Dynamics need to be represented as ODE, possibly constraining fidelity of model used in optimization & No constraints on dynamics representation.  Agent can learn in a high fidelity simulator (i.e., Navier-Stokes modeling of aerodynamics) \\
       Open Loop (requires a controller to track the optimal trajectory) & Closed Loop (Integrated guidance and control) \\
      Output feedback (co-optimization of state estimation and guidance law) an open problem for non-linear systems & Can learn from raw sensor outputs allowing fully integrated GNC (pixels to actuator commands). Can learn to compensate for sensor distortion.\\
      Requires full state feedback &   Does not require full state feedback \\
       Elegantly handles state constraints & State constraints handled either via large negative rewards and episode termination or more recently, modification of policy gradient algorithm. Control constraints straightforward to implement \\
      Deterministic Optimization & Stochastic Optimization, learning does not converge every time, may need to run multiple policy optimizations\\
   \end{tabular}
\end{table}

In this work, we  implement meta-RL using proximal policy optimization (PPO) \cite{schulman2017proximal} with both the policy and value function implementing recurrent layers in their networks.  To understand how recurrent layers result in an adaptive agent, consider that given some ground truth agent position, velocity, attitude, and rotational velocity $\mathbf{x}_{t}$, and action vector $\mathbf{u}_{t}$ output by the agent's policy, the next state $\mathbf{x}_{t+1}$ and observation $\mathbf{o}_{t+1}$ depends not only on $\mathbf{x}_{t}$ and $\mathbf{u}_{t}$, but also on the ground truth agent mass, inertia tensor, target maneuvers, and external forces acting on the agent. Consequently, during training, the hidden state of a network's recurrent network evolves differently depending on the observed sequence of observations from the environment and actions output by the policy. Specifically, the trained policy's hidden state captures unobserved (potentially time-varying) information such as external forces that are useful in minimizing the cost function. In contrast, a non-recurrent policy (which we will refer to as an MLP policy), which does not maintain a persistent hidden state vector, can only optimize using a set of current observations, actions, and advantages, and will tend to under-perform a recurrent policy on tasks with randomized dynamics, although as we have shown in \cite{gaudet2018deep}, training with parameter uncertainty can give good results using an MLP policy, provided the parameter uncertainty is not too extreme.  After training, although the recurrent policy's network weights are frozen, the hidden state will continue to evolve in response to a sequence of observations and actions, thus making the policy adaptive.  In contrast, an MLP policy's behavior is fixed by the network parameters at test time.

The PPO algorithm used in this work  is a  policy gradient algorithm which has demonstrated state-of-the-art performance for many RL benchmark problem. PPO approximates the TRPO optimization process\cite{schulman2015trust} by accounting for the policy adjustment constraint with a clipped objective function. The objective function used with PPO can be expressed in terms of the probability ratio $p_{k}({\bm \theta})$ given by,
\begin{equation}
\label{eq:clipr} 
p_{k}({\bm \theta})=\frac{\pi_{{\bm \theta}}({\bf u}_{k}|{\bf o}_{k})}{\pi_{{\bm \theta}_\text{old}}({\bf u}_{k}|{\bf o}_{k})}
\end{equation}
where the PPO objective function is then as follows:
\begin{equation}
\label{eq:ppoloss}
J({\bm \theta})=\mathbb{E}_{p({\bm \tau})}\left[\mathrm{min}\left[p_{k}({\bm \theta}) , \mathrm{clip}(p_{k}({\bm \theta}) , 1-\epsilon, 1+\epsilon)\right]A^{\pi}_{\bf w}({\bf o}_{k},{\bf u}_{k})\right]
\end{equation}
This clipped objective function has been shown to maintain a bounded KL divergence with respect to the policy distributions between updates, which aids convergence by insuring that the policy does not change drastically between updates. Our implementation of PPO uses an approximation to the advantage function that is the difference between the empirical return and a state value function baseline, as shown in Equation \ref{eq:ppo_adv}:
%\begin{equation}
%\label{eq:ppo_adv_old}
%   A^{\pi}_{\bf w}({\bf o}_{k},{\bf u}_{k})=r_k({\bf o}_k,{\bf u}_k) +\gamma V_{\bf w}^{\pi}({\bf o}_{k+1})-V_{\bf w}^{\pi}({\bf o}_k)
%\end{equation}
\begin{equation}
\label{eq:ppo_adv}
    A^{\pi}_{\bf w}(\mathbf{x}_{k},\mathbf{u}_{k})=\left[\sum_{\ell=k}^{T}\gamma^{\ell-k}r(\bf o_{\ell},\bf u_{\ell})\right]-V_{\bf w}^{\pi}(\mathbf{x}_{k})
\end{equation}
Here the value function $V_{\bf w}^{\pi}$ is learned using the cost function given by
\begin{equation}
\label{eq:vf_ppo}
L(\mathbf{w})=\sum_{i=1}^{M}\left(V_{\mathbf{w}}^{\pi}({\bf o}_k^i)-\left[\sum_{\ell=k}^{T}\gamma^{\ell-k}r({\bf u}_{\ell}^i,{\bf o}_{\ell}^i)\right]\right)^2
\end{equation}
In practice, policy gradient algorithms update the policy using a batch of trajectories (roll-outs) collected by interaction with the environment. Each trajectory is associated with a single episode, with a sample from a trajectory collected at step $k$ consisting of observation ${\bf o}_{k}$, action ${\bf u}_{k}$, and reward $r_k({\bf o}_k,{\bf u}_k)$. Finally, gradient ascent is performed on ${\bm \theta}$ and gradient decent on ${\bf w}$ and update equations are given by
\begin{align}\label{loss}
{\bf w}^+&={\bf w}^--\beta_{{\bf w}}\nabla_{{\bf w}} \left. L({\bf w})\right|_{{\bf w}={\bf w}^-}\\
{\bm \theta}^+&={\bm \theta}^-+\beta_{{\bm \theta}} \left. \nabla_{\bm \theta}J\left({\bm \theta}\right)\right|_{{\bm \theta}={\bm \theta}^-}
\end{align}
where $\beta_{{\bf w}}$ and $\beta_{{\bm \theta}}$ are the learning rates for the value function, $V_{\bf w}^{\pi}\left({\bf o}_k\right)$, and policy, $\pi_{\bm \theta}\left({\bf u}_k|{\bf o}_k\right)$, respectively.

In our implementation, we dynamically adjust the clipping parameter $\epsilon$ to target a KL divergence between policy updates of 0.001. The policy and value function are learned concurrently, as the estimated value of a state is policy dependent. The policy uses a Multi-categorical policy distribution, where a separate observation conditional categorical distribution is maintained for each element of the action vector. Note that exploration in this case is conditioned on the observation, with the two logits associated with each element of the action vector determining how peaked the softmax distribution becomes for each action. Because the log probabilities are calculated using the logits, the degree of exploration automatically adapts during learning such that the objective function is maximized. 

\subsection{Guidance Law Optimization}

Our guidance law was motivated by recent experimental research into the method falcons use to intercept prey \cite{kane2014falcons}, where it was found that the falcons attempt to keep the prey centered at the shallow fovea, a region of high photo-receptor density off set approximately 9 degrees from the falcon's body frame axis that is kept aligned with its velocity vector. The falcon then maneuvers in a way that keeps the prey centered at this location while minimizing the optical flow field. We found that results improved if we targeted seeker angles at their value at the start of the homing phase.  Specifically, (see Eqs.~\ref{eq:seeker3} and \ref{eq:seeker4}) we define $\theta_{u_o}$ and $\theta_{v_o}$ as the seeker angles at the start of the homing phase, and define the error signals $e_u=\theta_u-\theta_{u_o}$ and $e_v=\theta_v-\theta_{v_o}$.  We also define the change in $\theta_{u}$ and $\theta_{v}$ over the 100 mS guidance cycle as $d\theta_u,d\theta_v$. The observation given to the agent is then as shown in Eq.~\eqref{eq:obs}.

\begin{equation}
    \label{eq:obs}
    \mathrm{obs} = \begin{bmatrix} e_u & e_v & d\theta_u & d\theta_v  \end{bmatrix} 
\end{equation}

The action space for the guidance policy is in $\mathbb{Z}^{k}$, where $k$ is the number of thrusters.  We use a multi-categorical policy distribution.  During optimization, the policy samples from this distribution, returning a value in $\mathbb{Z}^{k}$,  Each element of the agent action $\mathbf{u} \in {0,1}$ is used to index Table \ref{tab:thrusters}, where if the  action is 1, it is used to compute the body frame force and torque contributed by that thruster.   For testing and deployment, the sampling is turned off, and the action is just the argmax of the two logits across each element of the action vector. 

The policy and value functions are implemented using four layer neural networks with tanh activations on each hidden layer. Layer 2 for the policy and value function is a recurrent layer implemented using gated recurrent units \cite{chung2015gated}. The network architectures are as shown in Table \ref{tab:NN}, where $n_{\mathrm{hi}}$ is the number of units in layer $i$, $\mathrm{obs\_dim}$ is the observation dimension, and $\mathrm{act\_dim}$ is the action dimension. The policy and value functions are periodically updated during optimization after accumulating trajectory rollouts of 30 simulated episodes.

\begin{table}[h]
	\fontsize{10}{10}\selectfont
    \caption{Policy and Value Function network architecture}
   \label{tab:NN}
        \centering 
   \newcolumntype{R}{>{\raggedleft\arraybackslash}p{1.8cm}}
   \begin{tabular}{l | R | c | R | c } % Column formatting, 
      \hline 
       & \multicolumn{2}{c}{Policy Network}\vline & \multicolumn{2}{c}{Value Network}\\
       \hline
       Layer & \# units & activation & \# units & activation \\
       \hline
      hidden 1      & $10 * \mathrm{obs\_dim}$ & tanh & $10 * \mathrm{obs\_dim}$ & tanh \\
      hidden 2      & $\sqrt{n_{\mathrm{h1}} * n_{\mathrm{h3}}}$ & tanh & $\sqrt{n_{\mathrm{h1}} * n_{\mathrm{h3}}}$ & tanh\\
      hidden 3      & $10 * \mathrm{act\_dim}$ & tanh & 5 & tanh \\
      output        & $\mathrm{act\_dim}$ & linear & 1 & linear \\
      \hline
   \end{tabular}
\end{table}

The agent receives a terminal reward  if the miss distance is less than 50 cm at the end of an episode.  Because it is highly unlikely that the agent will experience these rewards through random exploration, we augment the reward function using shaping rewards \cite{ng2003shaping}. These shaping rewards are given to the agent at each timestep, and guide the agent's behavior in such a way that the agent will begin to experience the terminal reward. Specifically, the shaping rewards encourage behavior that  minimizes the angle errors $e_u, e_v$ and the angular rate of change $d\theta_u,d\theta_v$, as shown in Eq.~\eqref{eq:rew1}, where $\alpha$ , $\sigma_{e}$ and $\sigma_{d\theta}$ are hyperparameters. The multiplicative Gaussian form maximizes shaping rewards when both the angle errors and angular rate of change are minimized. 

\begin{subequations}
\begin{align}
    r_{\mathrm{shaping}} &= \mathrm{exp}\left(-\frac{\|\begin{bmatrix} e_u & e_v  \end{bmatrix}\|}{\sigma_{e}} - \frac{\|\begin{bmatrix} d\theta_u & d\theta_v  \end{bmatrix}\|}{\sigma_{d\theta}}\right)\label{eq:rew1}\\
    r_{\mathrm{terminal}} &=
    \begin{cases}
        10,& \text{if } \mathrm{miss} < 50\mathrm{cm}\\
        0,              & \text{otherwise}
    \end{cases} \label{eq:rew2}\\
    r &= \alpha r_{\mathrm{shaping}} + r_{\mathrm{terminal}}\label{eq:rew3}
\end{align}
\end{subequations}

\begin{figure}[h]
\begin{center}
\includegraphics[width=.75\linewidth]{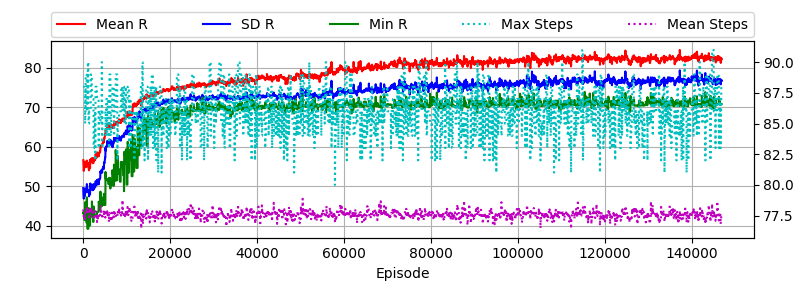}
\caption{Learning Curves: Rewards}
\label{fig:lc_rewards}
\end{center}
\end{figure}

\begin{figure}[h!]
\begin{center}
\includegraphics[width=.75\linewidth]{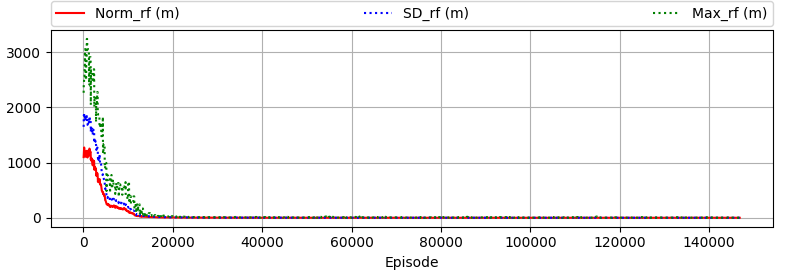}
\caption{Learning Curves: Rewards}
\label{fig:lc_miss}
\end{center}
\end{figure}

During optimization, the policy and value function are updated using rollouts collected over 30 episodes. An episode is terminated if one of the seeker angles $\theta_u, \theta_v$ exceeds the maximum field of view (135 degrees). Note that this termination condition covers the cases of successful intercepts and misses, and indirectly implements a field of view constraint.  We use the dual discount rate approach first suggested in \cite{gaudet2018deep}, with shaping rewards discounted by $\gamma_1=0.90$ and terminal rewards discounted by $\gamma_2=0.995$. Fig. \ref{fig:lc_rewards}  plots the reward statistics for the rewards received by the agent and number of steps per episode over the 30 episodes of rollouts used to update the policy and value function.  Note that a 100 step episode will have a duration of 10s.  Fig.  \ref{fig:lc_miss} gives statistics for miss distance during optimization, again with the statistics calculated over 30 episodes of rollouts. Here the "SD R" curve is the mean reward less one standard deviation. Note that the guidance policy was optimized assuming stabilized attitude, i.e., the missile's attitude remains fixed during the engagement.

We did experiment with several variations for the RL problem formulation, and found that including the bonus reward for a successful hit was crucial in obtaining good performance. Including the $d\theta_u$ and $d\theta_v$ terms is also important.  We did not test different policy and value function architectures or determine the effect of varying the terminal and shaping reward coefficients, and it is possible performance could be improved with a more extensive hyperparameter search. 

\section{Experiments}

\subsection{Augmented ZEM Guidance Law Benchmark}

In the following experiments, we compare the performance of the RL derived guidance law to that of the augmented zero-effort miss (ZEM) guidance law.  The augmented ZEM guidance law maps an estimate of the relative position and velocity between the missile and target and an estimate of the target acceleration to a commanded acceleration in the inertial reference frame, as shown in Eqs. \eqref{eq:zem1} through \eqref{eq:zem4}, where $N$ is optimally set to 3 for augmented ZEM. 

\begin{subequations}
\begin{align}
    \mathbf{ZEM} &= \mathbf{r}_\mathrm{TM} + \mathbf{v}_\mathrm{TM}t_{go} + \frac{1}{2}\mathbf{a}_{T}t_{go}^2\label{eq:zem1}\\
    v_c &= -\frac{\mathbf{r}_\mathrm{TM} \cdot \mathbf{v}_\mathrm{TM}}{\|\mathbf{r}_\mathrm{TM}\|}\label{eq:zem2}\\
    t_{go} &= \frac{\|\mathbf{r}_\mathrm{TM}\|}{v_c}\label{eq:zem3}\\
    \mathbf{a}_{\mathrm{com}} &= N \frac{\mathbf{ZEM}}{t_{go}^2}\label{eq:zem4}\\
\end{align}
\end{subequations}

We then convert the commanded acceleration to a body frame acceleration $\mathbf{a}_\mathrm{com}^B$ by projecting it onto the thruster model direction vectors in the first four rows of Table \ref{tab:thrusters}.  Pulsed  thrust is then achieved by turning on an engine when the corresponding element of $\mathbf{a}_\mathrm{com}^B$ exceeds 1/3 of the maximum acceleration (the ratio of maximum thrust to dry mass). We checked that performance is  close to that of acceleration limited augmented ZEM where the acceleration is directly applied to the missile (i.e., Eq.~\eqref{eq:TEQOMb}).  Our approach allows a closer comparison of the two guidance laws as we can use the same equations of motion.

\subsection{Comparison of RL guidance policy with Augmented ZEM}

In order to provide a good comparison with the augmented ZEM guidance law,  we assume that the missile's attitude is perfectly stabilized, i.e., the attitude remains unchanged from that at the start of the homing phase.  The RL policy and augmented ZEM policy are tested against the engagement scenario described in Section \ref{engagement}, where the values given in Table \ref{tab:ic} are randomized in each episode. The environment used for optimization is also used for testing. Randomizing each episode's engagement parameters insures that during testing, the agent experiences novel engagement scenarios not experienced during optimization.

Table \ref{tab:Comparison} tabulates the results from running 5000 simulations for the RL optimized policy and the augmented ZEM policy using randomized initial conditions with bounds given in Table \ref{tab:ic}. The RL policy exhibits better accuracy, despite the fact that the augmented ZEM policy is given the full engagement state (relative position, velocity, and target acceleration) as an observation, whereas the RL policy is only using seeker angles and their rate of change.  The RL policy is also more fuel efficient, which is important for exo-atmospheric intercepts where the interceptor will lose all control authority when its fuel is exhausted. We repeat this experiment, but without randomizing heading error and target acceleration (they are set to the maximum values) with the heading error at 5 degrees, attitude error at 5 degrees, and maximum target acceleration. Note that for these worst case values, the heading error and attitude error is still random, i.e., there are an infinite number of initial missile velocity vectors with an angle of 5 degrees with respect to the ideal missile velocity vector in $\mathbb{R}^3$. Results are shown in Table \ref{tab:Comparison_wc}.

\begin{table}[!ht]
	%\fontsize{9}9\selectfont
    \caption{RL / Augmented ZEM Comparison: Randomized Heading Error and Target Acceleration}
   \label{tab:Comparison}
        \centering 
   \begin{tabular}{l | r | r | r | r } % Column formatting, 
      \hline
      Parameter & \multicolumn{2}{c}{Miss (cm)}\vline & \multicolumn{2}{c}{Fuel (kg)}\\
      \hline
      Value & $<$ 100 cm (\%) & $<$ 50 cm (\%) & $\mu$ &  $\sigma$\\
      \hline
      Aug. ZEM (N=3)  & 97 & 45 & 9.4 & 3.7 \\
      RL   &  99 & 68 & 7.8  & 2.8
   \end{tabular}
\end{table}

\begin{table}[!ht]
	\fontsize{10}{10}\selectfont
    \caption{RL / Augmented ZEM Comparison: Maximum Heading Error and Target Acceleration}
   \label{tab:Comparison_wc}
        \centering 
   \begin{tabular}{l | r | r | r | r } % Column formatting, 
      \hline
      Parameter & \multicolumn{2}{c}{Miss (cm)}\vline & \multicolumn{2}{c}{Fuel (kg)}\\
      \hline
      Value & $<$ 100 cm (\%) & $<$ 50 cm (\%) & $\mu$ &  $\sigma$\\
      \hline
      Aug. ZEM (N=3) & 83 &  28 & 17.5 & 2.1 \\
      RL   &  95 & 53 & 16.1  & 2.1
   \end{tabular}
\end{table}

\begin{figure}[h]
\begin{center}
\includegraphics[width=.75\linewidth]{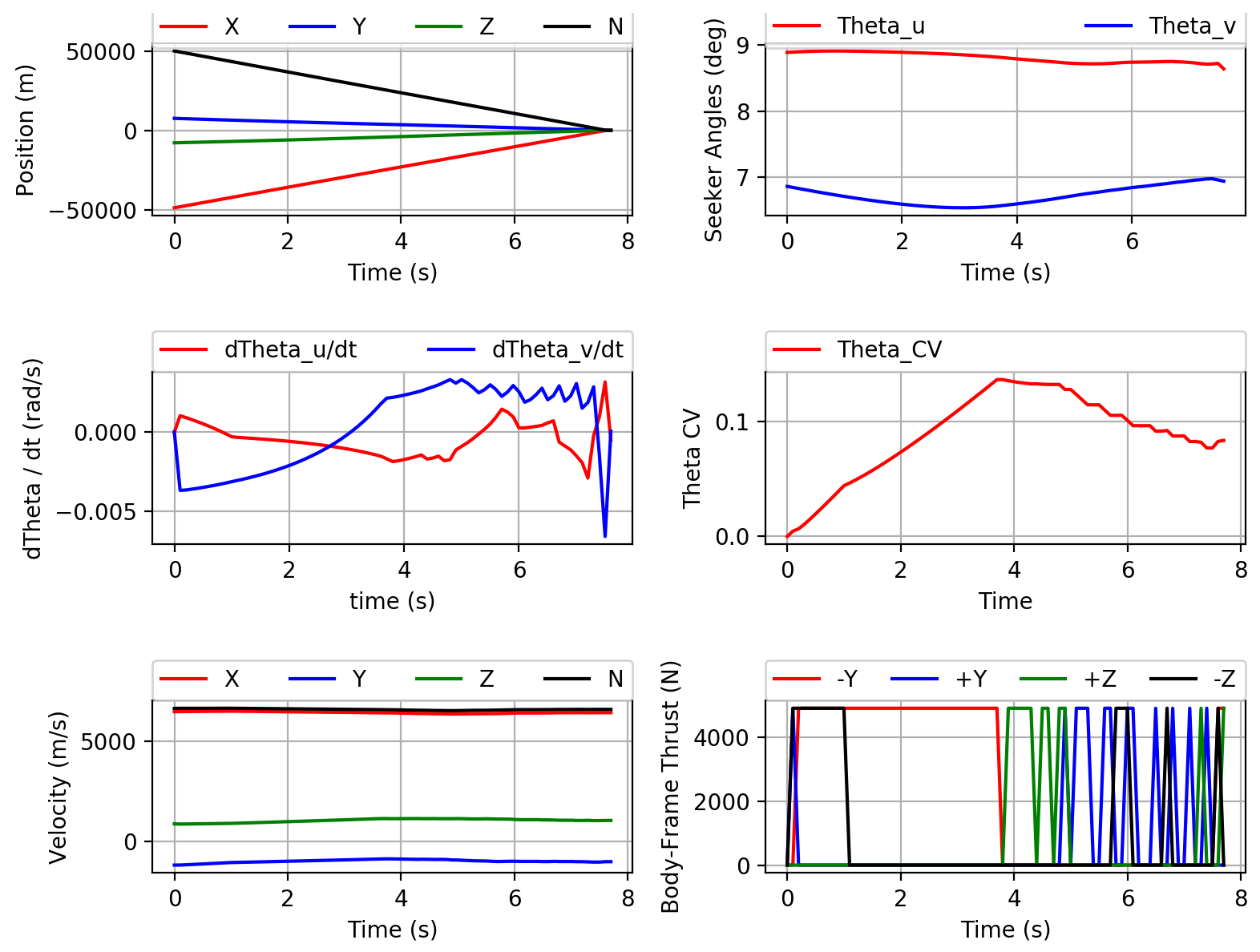}
\caption{Sample RL Trajectory}
\label{fig:rl_traj}
\end{center}
\end{figure}

\begin{figure}[H]
\begin{center}
\includegraphics[width=.75\linewidth]{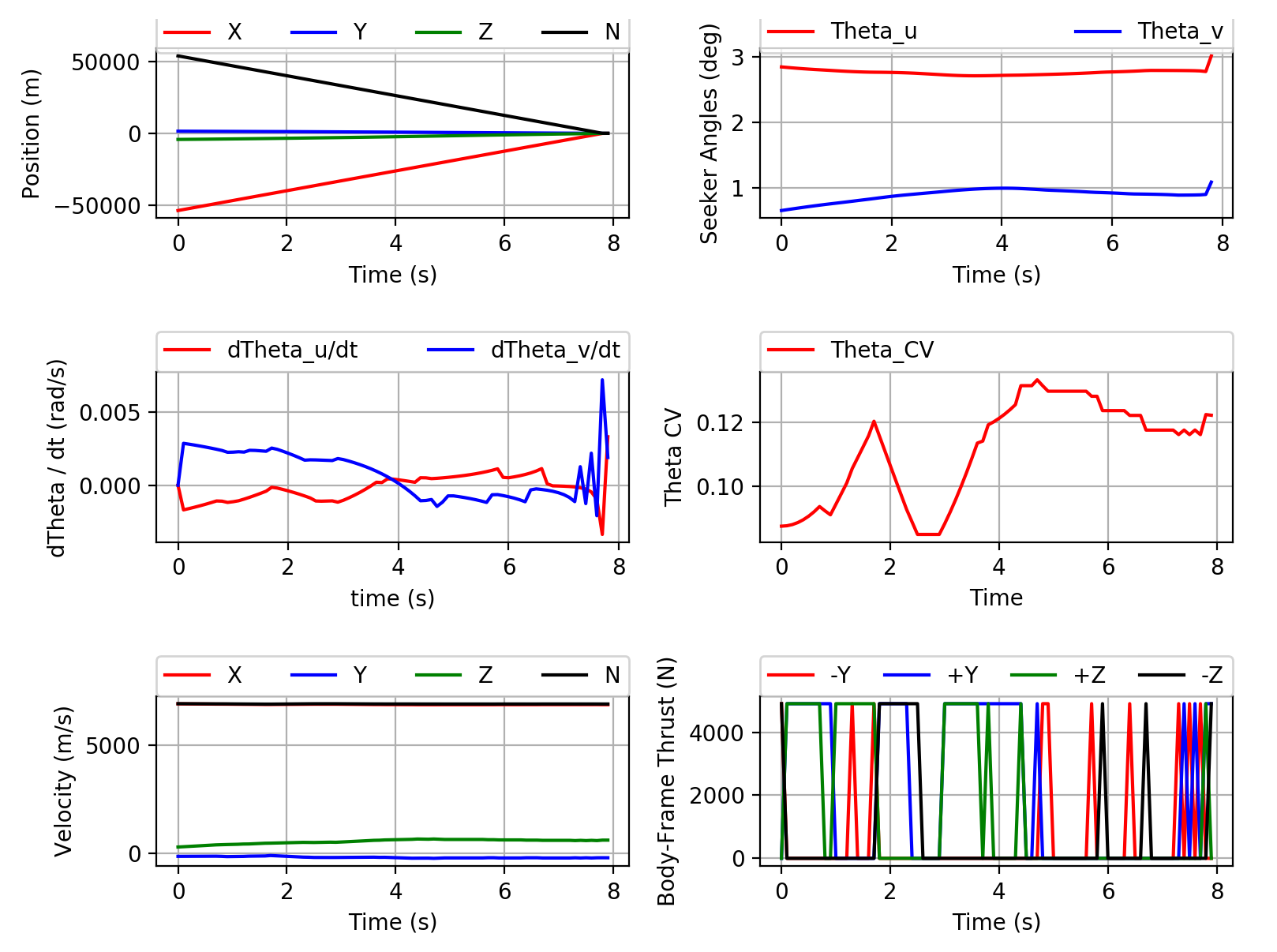}
\caption{Sample Augmented ZEM Trajectory}
\label{fig:zem_traj}
\end{center}
\end{figure}

Figs. ~\ref{fig:rl_traj} and \ref{fig:zem_traj} illustrate a trajectory corresponding to randomly generated (and different in each case) initial conditions.  Position is the missile's position in a target centered reference frame. In the seeker angle and angle rate of change plots, Theta\_u and Theta\_v correspond to $\theta_u$ and $\theta_v$ respectively. Although the augmented ZEM policy does not use the seeker, the simulator still instantiates a seeker so that we can view the seeker angles and their rate of change during the engagement. The subplot labeled "Theta\_CV" plots the angle between the missile's velocity vector and the missile's body frame x-axis. Note that even for the case where Theta\_CV starts at zero, missile divert thrusts will in general cause a misalignment between these vectors. Although in theory these could be corrected using attitude control thrusters, this would require knowledge of the missile's velocity vector, which in this work we assume is unknown.   At any rate, the effect on performance is minimal.  It appears that the RL policy is doing what we would hope, expending most of the control effort early in the engagement.

\subsection{Policy Generalization to Novel Engagement Scenarios}
 In order to  test the ability of the optimized policy to generalize to engagement parameters outside the range experienced during optimization, we extend the range of several engagement parameters as shown in Table \ref{tab:ext_ic}, with the other engagement parameters unchanged from Table \ref{tab:ic}.  We found accuracy was unchanged, with fuel consumption increasing by approximately 1kg. Next we test the RL policy on a novel target maneuver not seen during optimization, a  barrel roll maneuver \cite{zarchan2012tactical:3} with randomized weave period ranging from 1s to 5s, and the magnitude of target acceleration constant at the target's maximum acceleration capability. This ended up being a more difficult maneuver for both the RL and augmented ZEM policies. The results are shown in Table \ref{tab:Comparison_wc_barrel_roll}, with augmented ZEM results given for comparison. It is possible that the performance of the RL policy could be improved in this case by optimizing over a mix of bang-bang and barrel roll maneuvers. Intuitively, we expect good generalization if the novel engagement scenarios result in seeker angles and their rates of change staying within the ranges experienced during optimization. Finally we compare performance between the RL policy and the augmented ZEM policy when the heading error is increased to 6 degrees. Here we do not randomize the heading error and target acceleration (they are set to the maximum values). Otherwise the engagement parameters are as shown in Table \ref{tab:ic}. Since we chose the engagement scenario parameters in Table \ref{tab:ic} to create an engagement at the limit of the augmented ZEM policy's capability, we expect the performance for both policies to deteriorate as compared to the results shown in Table \ref{tab:Comparison_wc}. The results are shown in Table \ref{tab:Comparison_HE}. It is possible that we could have improved the RL policy results by optimizing over an extended range of heading errors.

 \begin{table}[H]
	\fontsize{10}{10}\selectfont
    \caption{Extended Initial Conditions}
   \label{tab:ext_ic}
        \centering 
   \begin{tabular}{l | r | r } % Column formatting, 
      \hline
      Parameter & min & max \\
      \hline
      Range (km) & 50 & 75\\
      Missile Velocity Magnitude (m/s) & 3000 & 3500 \\
      Target Position angle $\theta$ (degrees) & -20 & 20 \\
      Target Position angle $\phi$ (degrees) & -20 & 20 \\
      Target Velocity Magnitude (m/s) & 3000 & 4000 \\
      Target Velocity angle $\beta$ (degrees) & -15 & 15 \\
      Target Velocity angle $\alpha$ (degrees) & -15 & 15 \\
   \end{tabular}
\end{table}
 
\begin{table}[!h]
	\fontsize{10}{10}\selectfont
    \caption{RL / Augmented ZEM Comparison for Barrel Roll Target Maneuver}
   \label{tab:Comparison_wc_barrel_roll}
        \centering 
   \begin{tabular}{l | r | r | r | r } % Column formatting, 
      \hline
      Parameter & \multicolumn{2}{c}{Miss (cm)}\vline & \multicolumn{2}{c}{Fuel (kg)}\\
      \hline
      Value & $<$ 100 cm (\%) & $<$ 50 cm (\%) & $\mu$ &  $\sigma$\\
      \hline
      Aug. ZEM (N=3)  & 58 & 19 & 21.8 & 3.2 \\
      RL   &  59  & 25 & 13.5  & 3.5 \\
   \end{tabular}
\end{table}

\begin{table}[!h]
	\fontsize{10}{10}\selectfont
    \caption{RL / Augmented ZEM Comparison for 6 degree Heading Error}
   \label{tab:Comparison_HE}
        \centering 
   \begin{tabular}{l | r | r | r | r } % Column formatting, 
      \hline
      Parameter & \multicolumn{2}{c}{Miss (cm)}\vline & \multicolumn{2}{c}{Fuel (kg)}\\
      \hline
      Value & $<$ 100 cm (\%) & $<$ 50 cm (\%) & $\mu$ &  $\sigma$\\
      \hline
      Aug. ZEM (N=3)  &  79 & 28 & 18.8 & 1.97\\
      RL   &  93 & 49 & 16.1 & 2.1 \\
   \end{tabular}
\end{table}

\section{Conclusion}

We have demonstrated that it is possible to formulate a missile guidance problem in the RL framework that maps seeker angles and their rate of change directly to actuator commands. The resulting guidance policy has performance close to that of state of the art methods such as augmented ZEM with access to the ground truth engagement state.  We show that the guidance law generalizes well to novel engagement scenarios not experienced during optimization, specifically an extended initial condition range, novel target maneuvers, and increased heading error. The guidance law developed in this work is particularly applicable to passive seekers, which are not capable of measuring range or range rate. To date it has not been possible to formulate a guidance law using angle-only measurements using the optimal control framework (which typically requires full state feedback), and estimating the full engagement state from angle-only measurements is an open problem.  Further, optimizing a guidance law in a framework allowing the specification of a cost (reward) function and constraints has the potential to improve performance as compared to the state of the art systems using passive seekers. We therefore suggest that the RL framework is a good candidate for improving the performance of guidance systems compatible with passive seekers. Future work might include endo-atmospheric intercepts, more realistic engagement scenarios, higher fidelity dynamics, and applying RL to target discrimination.

\bibliographystyle{AAS_publication}   % Number the references.
\bibliography{references.bib}   % Use references.bib to resolve the labels.

\end{document}